\documentclass[10pt, conference, compsocconf]{IEEEtran}

\let\labelindent\relax
\usepackage{hyperref, pdfcomment}
\usepackage{url}
\usepackage{enumitem}
\usepackage{graphicx}
\usepackage{subcaption}
\usepackage{multirow}

\IEEEoverridecommandlockouts

\begin{document}

\title{BigExcel: A Web-Based Framework for Exploring Big Data in Social Sciences}

\author{\IEEEauthorblockN{Muhammed Asif Saleem, Blesson Varghese and Adam Barker}
\IEEEauthorblockA{School of Computer Science, University of St Andrews\\
St Andrews, Fife, UK KY16 9SX\\
Email: \{mas23, varghese, adam.barker\}@st-andrews.ac.uk}
}

\maketitle

\begin{abstract}
This paper argues that there are three fundamental challenges that need to be overcome in order to foster the adoption of big data technologies in non-computer science related disciplines: addressing issues of accessibility of such technologies for non-computer scientists, supporting the ad hoc exploration of large data sets with minimal effort and the availability of lightweight web-based frameworks for quick and easy analytics. In this paper, we address the above three challenges through the development of `BigExcel', a three tier web-based framework for exploring big data to facilitate the management of user interactions with large data sets, the construction of queries to explore the data set and the management of the infrastructure. The feasibility of BigExcel is demonstrated through two Yahoo Sandbox datasets. The first dataset is the Yahoo Buzz Score data set we use for quantitatively predicting trending technologies and the second is the Yahoo n-gram corpus we use for qualitatively inferring the coverage of important events. A demonstration of the BigExcel framework and source code is available at \url{http://bigdata.cs.st-andrews.ac.uk/projects/bigexcel-exploring-big-data-for-social-sciences/}.  
\end{abstract}


\IEEEpeerreviewmaketitle

\section{Introduction}
\label{introduction}

The era of `big data' \cite{definition} has transformed the data analysis landscape over the last few years\footnote{\url{http://archive.wired.com/science/discoveries/magazine/16-07/pb_theory}}. Transformative technologies, such as on demand cloud computing infrastructure, and programming paradigms such as Hadoop \cite{hadoop} and Hive \cite{hive-1} facilitate applications to perform large-scale data analysis. However, we argue that there are three fundamental challenges, which need to be addressed in order to allow domain experts without a Computer Science background to benefit from such technologies. 

\subsubsection*{Challenge 1 - Limited Accessibility of Big Data Tools}
While big data technologies such as Hadoop provide the tools to facilitate the analysis of large-scale data, there is significant research that still needs to be performed in order to make these tools accessible to the wider community\footnote{\url{http://www.wired.com/2013/05/the-importance-of-making-big-data-\\accessible-to-non-data-scientists/}}. In the current state-of-the-art, there is a gap between big data technologies and the end user\footnote{\url{http://chronicle.com/article/Recent-Big-Data-Struggles-Are/145625/}}; an in-depth knowledge of the technologies for setting up and maintaining hardware along with programming skills for accessing Application Programming Interfaces (API) are required for performing any meaningful analysis \cite{bigdata-1}. For example, the use of Hadoop or Hive still requires detailed knowledge of parallel programming using the MapReduce paradigm \cite{mapreduce-1, mapred-2}. The research reported in this paper aims toward bridging this gap by addressing how big data technologies can be made accessible to expert practitioners who are non-computer scientists, such as social scientists, for performing big data analytics. 

\subsubsection*{Challenge 2 - Lack of Exploratory Tools for Big Data} 
From discussions with social scientists, we understand that a typical working pattern with a data set or variety of data sets involves the exploration of data by performing a quick analysis first in order to infer whether the data is relevant or of interest before performing more complex analysis\footnote{\url{http://harvardmagazine.com/2014/03/why-big-data-is-a-big-deal}}. For example, given a large dataset related to newspaper articles from 1885--1975, how can a social scientist quickly explore the dataset to find which newspapers published articles related to an event in 1925? Such exploration needs to be performed with minimal effort without having to undertake large programming tasks. In this paper, we present how data exploration can be performed with minimal effort. 

\subsubsection*{Challenge 3 - Lack of Lightweight Big Data Tools}
Research efforts have been made towards the development of tools that can facilitate big data analytics. For example, the IBM InfoSphere \cite{ibminfosphere} and Tableau software \cite{tableau}. These tools provide fully fledged functionalities and are suitable for those who receive training for using them proficiently. Such tools are not suited and required for exploring data and performing simple analytical operations on data. In this research, we aim to develop a lightweight tool that provides essential functionalities to explore data and perform basic analytical operations. 

In this paper, we address the above three challenges related to the accessibility of big data technologies for non-computer scientists, the exploration of large data sets with minimal efforts and the availability of lightweight frameworks for ad hoc analytics by developing a web-based framework, we refer to as BigExcel, for exploring big data. The three tier framework is situated between a user and the underlying infrastructure to facilitate the management of user interactions with large data sets, the construction of queries to explore the data set and the management of the infrastructure. The feasibility of BigExcel is validated on two case studies using Yahoo Sandbox \cite{yahoosandbox} datasets.   

The contributions of the research presented in this paper are (i) the development of a lightweight web-based framework for exploring and performing analytics in an ad hoc manner on big data, (ii) facilitating simplified access to big data technologies for social scientists by developing an interactive layer between the user and the underlying hardware infrastructure, (iii) the abstraction of technicalities related to the setup, maintenance and execution of big data and the queries associated with large datasets on big data technologies such as Hive and Hadoop Distributed File System (HDFS), and (iv) the demonstration of the framework on two datasets provided by Yahoo Sandbox.  

The remainder of this paper is organised as follows. Section \ref{framework} presents the web-based framework developed in this research to facilitate the exploration and visualisation of big data. Section \ref{casestudies} presents two case studies which are used to validate the feasibility of the framework and demonstrate how typical datasets can be explored. The first case study is based on quantitative prediction of trending technologies and the second case study is based on a qualitative analysis of n-grams from news oriented websites correspond with real events. Section \ref{conclusions} concludes this paper by presenting a short discussion on future work. 

\section{Framework}
\label{framework}

\begin{figure} [!ht]
	\centering
	\includegraphics[width = 0.46\textwidth]{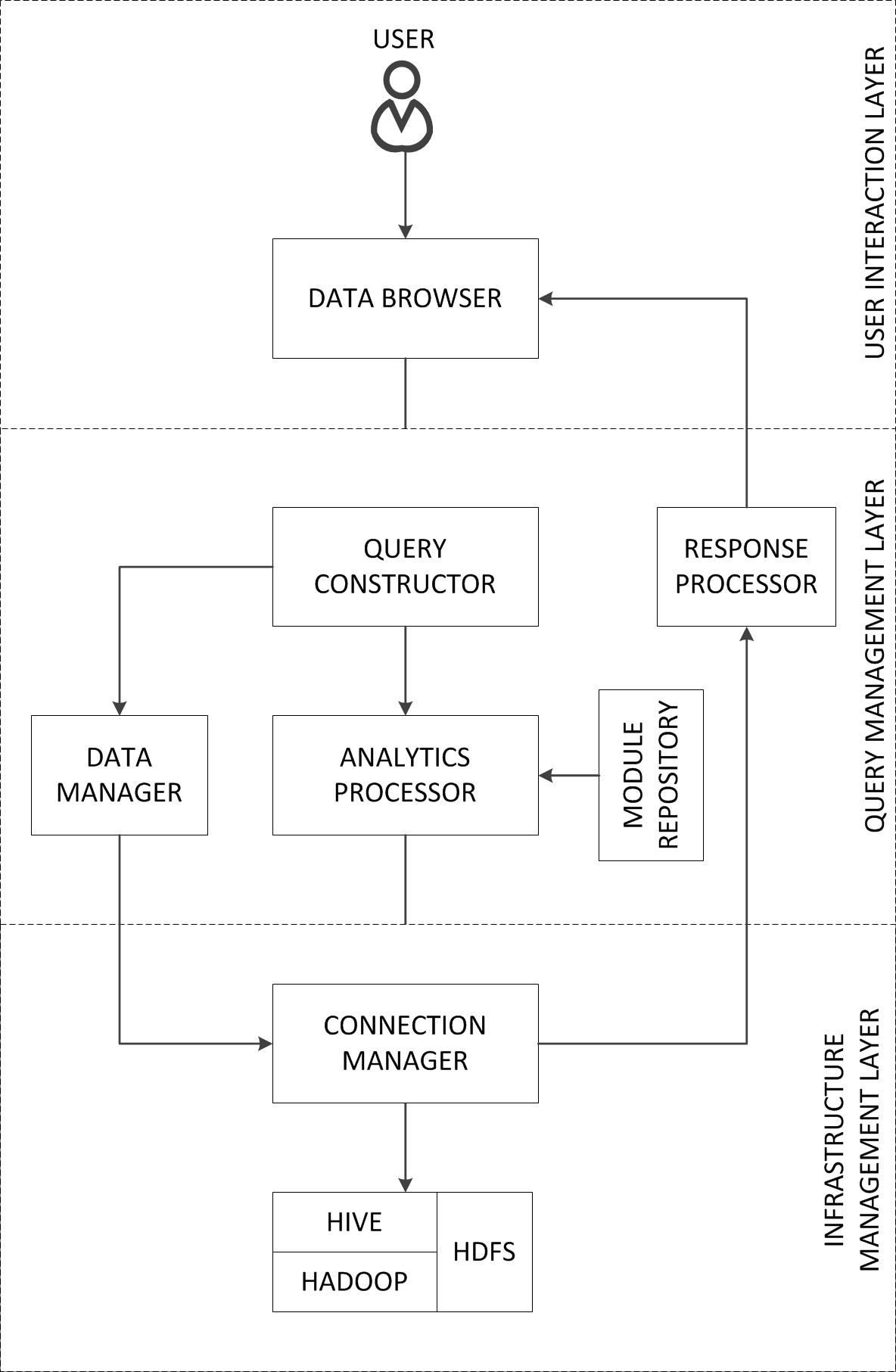}
	\caption{The BigExcel web-based framework for exploring big data in real-time}
	\label{figure0}
\end{figure}

BigExcel: a web-based framework for exploring big data in real-time is illustrated in Figure \ref{figure0}. BigExcel consists of three layers; the first layer is responsible for handling user interactions, the second layer manages queries that express how data needs to be handled, and the third layer is related to the management of underlying infrastructure.

\subsection{User Interaction Layer}
This layer handles all user interactions in the BigExcel framework, such as requesting a specific analytical operation on data and the display of the operation in a meaningful way. This layer consists of one module, referred to as the \texttt{Data Browser} which is built using RichFaces \cite{richfaces}, a framework for developing UIs and supported by jQuery \cite{jquery} for data validation and navigation and JavaServer Faces (JSF) \cite{jsf} that facilitates data communication with the next layer. The browser enables a user to submit a request for managing data (for example, loading or viewing data) and for performing analytics on this data. Using RESTful (REpresentational State Transfer) Web Services \cite{restful-1} the Application Programming Interface (API) provided by the next layer is used by the browser. 

\subsection{Query Management Layer}
This layer receives data from the user interaction layer through JSF and is responsible for constructing queries, attaching relevant modules to the queries for analytics, and handling the responses from the next layer. Five modules are used in this layer, and they are as follows:  

\subsubsection{\normalfont \texttt{Query Constructor}}
User requests obtained from the browser are converted to queries that can executed on Hive \cite{hive-1}. Two types of queries are constructed; the first are data management queries which are handled by the \texttt{Data Manager} and the second is analytics queries handled by the \texttt{Analytics Processor}. 

\subsubsection{\normalfont \texttt{Data Manager}}
The data management queries are handled by this module. They are related to transferring data to the Hadoop Distributed File System (HDFS) \cite{hdfs-1} and loading data from HDFS to Hive. 

\subsubsection{\normalfont \texttt{Analytics Processor}}
Three types of analytics queries are handled by this module: (a) the set of queries which make use of built-in modules, such as sum, average and standard deviation, that apply to rows or columns of structured big data and execute on Hive, (b) the set of queries which can incorporate user defined business logic in the form of modules that can be execute on Hive, and (c) the set of queries which incorporates user defined logic in the form of MapReduce \cite{mapreduce-1} modules that can be executed on Hive or Hadoop. 

\subsubsection{\normalfont \texttt{Module Repository}}
It may not be always possible to derive the business logic for an analytical operation using the combination of built-in modules. User defined modules are therefore required to incorporate business logic into queries. The module repository, which is a database of user defined modules that are incorporated into queries by the \texttt{Analytics Processor} facilitates this functionality.

\subsubsection{\normalfont \texttt{Response Processor}}
This module prunes data for presenting to the user in the data browser. In the case of big data, the dynamic pagination (or lazy loading) technique is used for facilitating real-time viewing of data on the browser. In this technique, chunks of data are displayed on-demand. The \texttt{Response Processor} also employs chart APIs (for example, Google Charts \cite{googlecharts}) for presenting data on the browser.  

\subsection{Infrastructure Management Layer}
This layer handles the communication between the query management layer and the underlying infrastructure. The key module that facilitates this is the \texttt{Connection Manager} which follows a multi-threaded model for handling concurrent users. The Amazon Web Services (AWS) SDK \cite{awssdk} is used to connect to the cloud. 

\section{Case Studies}
\label{casestudies}

The BigExcel framework presented in the previous section was tested using two case studies employing data sets from Yahoo Sandbox. The first case study uses marketing data and the second case study uses language data. The case studies are used to demonstrate the feasibility of the framework. The source code of the framework along with a video illustrating the functionality of the framework is available at \url{http://bigdata.cs.st-andrews.ac.uk/projects/bigexcel-exploring-big-data-for-social-sciences/}.

\subsection{Case Study 1: Predicting Market Trends Based on Search Volumes}
In this case study, the data set employed is the `Yahoo! Buzz Game Transactions with Buzz Scores'\footnote{\url{http://webscope.sandbox.yahoo.com/catalog.php?datatype=a}}, version 1.0, which is nearly 90 MB (uncompressed data). The data contains information on various trending technologies and the volume of searches associated with the technology, which is indicated as a buzz score, for a period from April 1, 2005 to July 31, 2005. A higher buzz score for a technology indicates a larger volume of search for that technology. In the dataset, technology is grouped into items, such as browsers, TVs, e-books, video games, social networks, photo organisers, and GPS/map related. Each technology consists of products, for example, browsers contain sub-items such as Internet Explorer, Safari, Mozilla, and Firefox. The data is relevant to a market in which stocks in various technologies are bought and sold based on current buzz scores. Hence, it is important to understand the fluctuations of the buzz score for making purchasing and selling decisions. 

We use this data in the BigExcel framework for (i) predicting hourly, daily and weekly patterns of buzz scores and (ii) for comparing the prediction against the actual buzz scores to evaluate whether prediction is within a reasonable order of magnitude accuracy. 

The dataset is obtained in text (.txt) format. The framework converts the .txt format to comma separated file (.csv) format and the .csv file is then loaded into the Hive warehouse. A user can then explore the data in BigExcel by plugging modules which are stored in the \texttt{Module Repository}.    

\subsubsection{Hourly Analysis}
\begin{figure*}
	\centering
	\begin{subfigure} {\textwidth}
		\centering
  		\includegraphics[width=\textwidth]{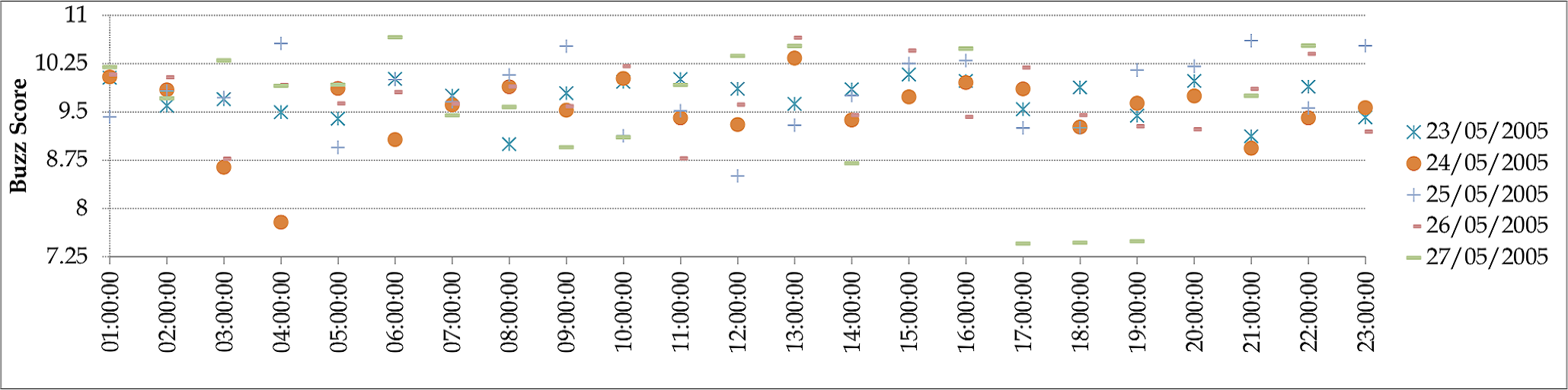}
  		\caption{23 - 27 May 2005}
  		\label{figure1a}
	\end{subfigure} \hfill
	\begin{subfigure} {\textwidth}
		\centering
  		\includegraphics[width=\textwidth]{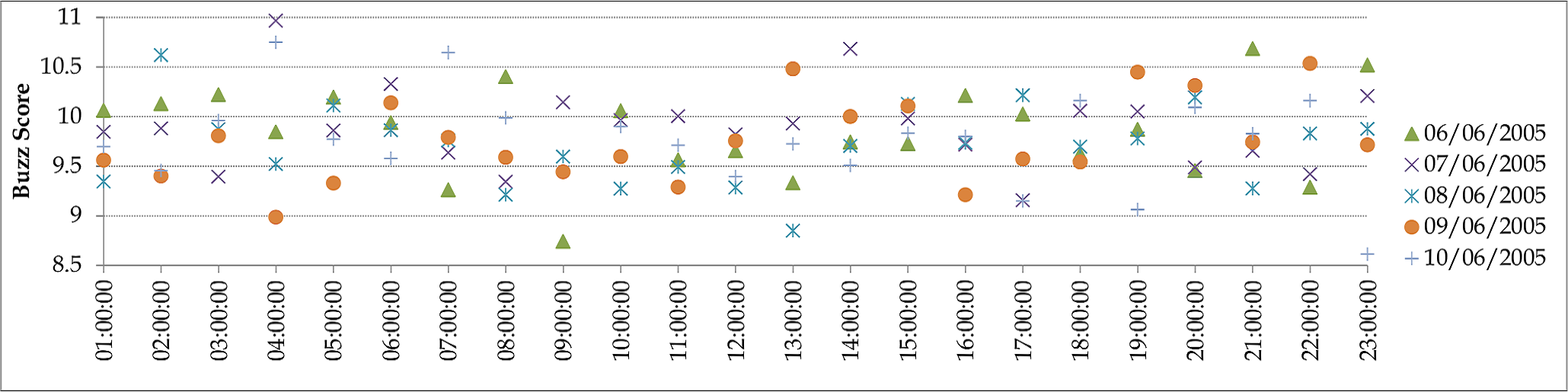}
  		\caption{06 - 10 June 2005}
  		\label{figure1b}
	\end{subfigure} 
	\caption{Buzz scores of video games on an hourly basis for a working week in May and June}
	\label{figure1}
\end{figure*}

An hourly analysis of the buzz scores of video games as shown in Figure \ref{figure1} for two working weeks, the first from 23/05/2005 to 27/05/2008 and the second from 06/06/2005 to 10/06/2005 was performed. Such an analysis can capture the hourly trends which may be useful to make observations of obvious patterns that might emerge. The hourly buzz scores are calculated by aggregating all the buzz scores related to video games in that hour. In Figure \ref{figure1}, for example, the buzz score shown for 09:00:00 for a given day is the average of all the buzz scores from 09:00:00 to 09:59:59 for that day.

To perform the above, the user provides as input the table name (\texttt{Yahoo\_Buzz\_Scores}), column names (\texttt{date}, \texttt{time} and \texttt{buzz\_score}), product type (\texttt{e-books} and \texttt{buzz\_score}) and the name of the module required for analysis (\texttt{hourly\_analysis}) through the data browser. The \texttt{Query Constructor} then assembles the following query:

\begin{verbatim}
SELECT TRANSFORM(date, time, buzz_score) 
USING 'hourly_analysis' 
FROM Yahoo_Buzz_Scores 
WHERE product='EBOOKS' 
AND date >= 2005-05-23 
AND date <=2005-05-27;
\end{verbatim}

The \texttt{Analytics Processor} embeds the \texttt{hourly\_analysis} module from the \texttt{Module Repository} into the query and the query is executed on Hive. The \texttt{hourly\_analysis} module includes the logic for averaging the buzz scores in a one hour period to generate an aggregate buzz score for the hour. The output is parsed by the \texttt{Response Processor} to display the information in a meaningful way, for example, to generate data shown in Figure \ref{figure1}. 

Immediate observations from Figure \ref{figure1} include the cluster of buzz scores for the week in both May and June at 1 am, 10 am, 11 am, 3 pm, 6 pm and 11 pm. A consistent occurrence of clusters across different weeks can be used as a pointer to predict with some certainty what the buzz scores may be for those time periods; this information will be useful while trading stocks based on buzz scores. Another observation includes the low buzz scores for both weeks on Wednesdays at 12 noon and 1 pm.   

\subsubsection{Daily Analysis}
\begin{figure*} [!ht]
	\centering
	\includegraphics[width = \textwidth]{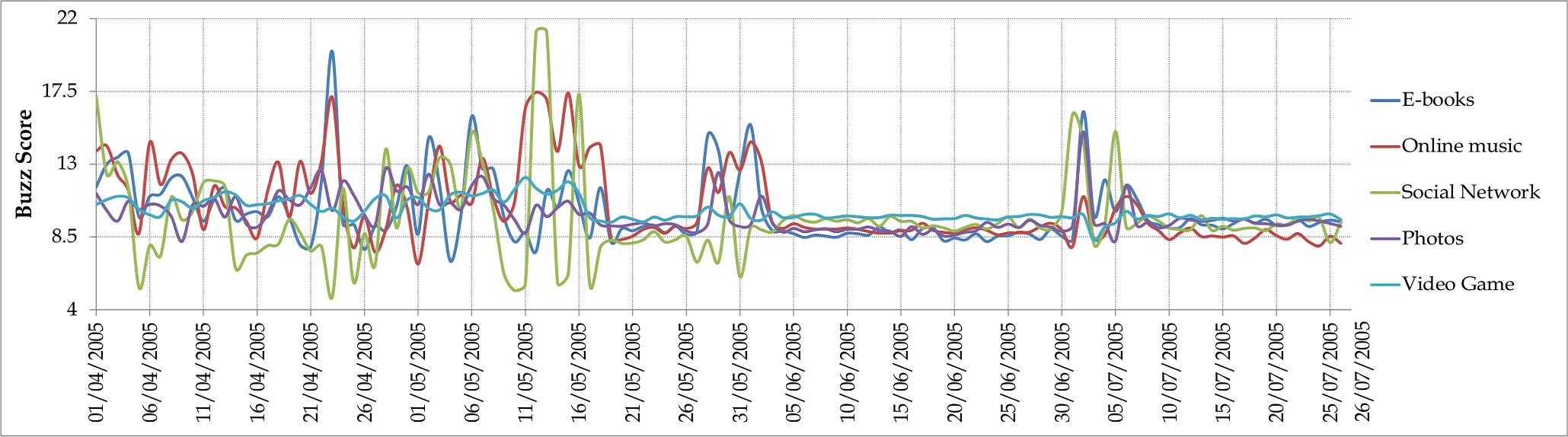}
	\caption{Daily aggregated buzz scores of technologies from April 1, 2005 to July 31, 2005}
	\label{figure2}
\end{figure*}

Figure \ref{figure2} shows the actual buzz scores of e-books, online music, social network, photo organisers and video games for each day in the period from April 1, 2005 to July 31, 2005. A user defined module that is stored in the \texttt{Module Repository} of the BigExcel framework is created to facilitate the computation of the buzz scores for each day. The buzz scores of one hour intervals of a day are averaged to aggregate the buzz score for a day. 

To perform the above, the user provides as input the table name (\texttt{Yahoo\_Buzz\_Scores}), column names (\texttt{date}, \texttt{buzz\_score}, products and the name of the module that for analysis (\texttt{hourly\_analysis}) through the data browser. The \texttt{Query Constructor} then assembles the following query:

\begin{verbatim}
SELECT TRANSFORM(date, buzz_score)
USING 'daily_analysis' 
FROM Yahoo_Buzz_Scores
WHERE product IN ('ONLNMUSIC','EBOOKS', 
	'VGAME', 'SOCNETS', 'PHOTO')
AND date>='2005-04-01' 
AND date <='2005-07-26'
\end{verbatim}

The \texttt{daily\_analysis} module is embedded from the \texttt{Module Repository} into the query by the \texttt{Analytics Processor}. The module includes the logic for averaging the hourly buzz scores in a given day to generate an aggregate buzz score for the day. The output is parsed by the \texttt{Response Processor} to display the information in a meaningful way, for example, to generate data that represents Figure \ref{figure2}. 

We incorporate two techniques for predicting the buzz score of a given day stored in the \texttt{daily\_prediction} module. In the first technique, the buzz scores of the preceding $n$ days or of the same week day in a sample of $n$ weeks are aggregated by averaging to predict the buzz score of a given day. For example, if the preceding $n$ days are taken into account, then to predict the buzz score of video games on July 23, 2005, the buzz scores from July 8 to July 22 are employed if 14 preceding days are considered, or the buzz scores from June 24 to July 22 are employed if 28 preceding days are used. If a sample of $n$ weeks are used, then to predict the buzz score of e-books on Saturday, July 23, 2005, the buzz scores on Saturdays, June 18, June 25, July 9 and July 16 are used if a four week sample is taken into account, or the buzz scores on Saturdays, May 14, June 4, June 18, June 25, July 2, July 9 and July 16 are used if an eight week sample is taken into account. We refer to this technique as `Extrapolation-based Prediction (EP)' with preceding $n$ days or a sample of $n$ weeks. 

In the second technique, the buzz score of the preceding $n$ days or of the same week day in a sample of $n$ weeks are aggregated by linear regression to predict the buzz score of a given day. This technique is different from the first and second techniques in that linear regression is applied on the buzz scores for predicting rather than by averaging. The data points used for the regression remains the same as in the former techniques. We refer to this technique as `Regression-based Prediction (RP)' with preceding $n$ days or a sample of $n$ weeks.

\begin{figure}
	\centering
	\begin{subfigure} {0.49\linewidth}
		\centering
  		\includegraphics[width=\linewidth]{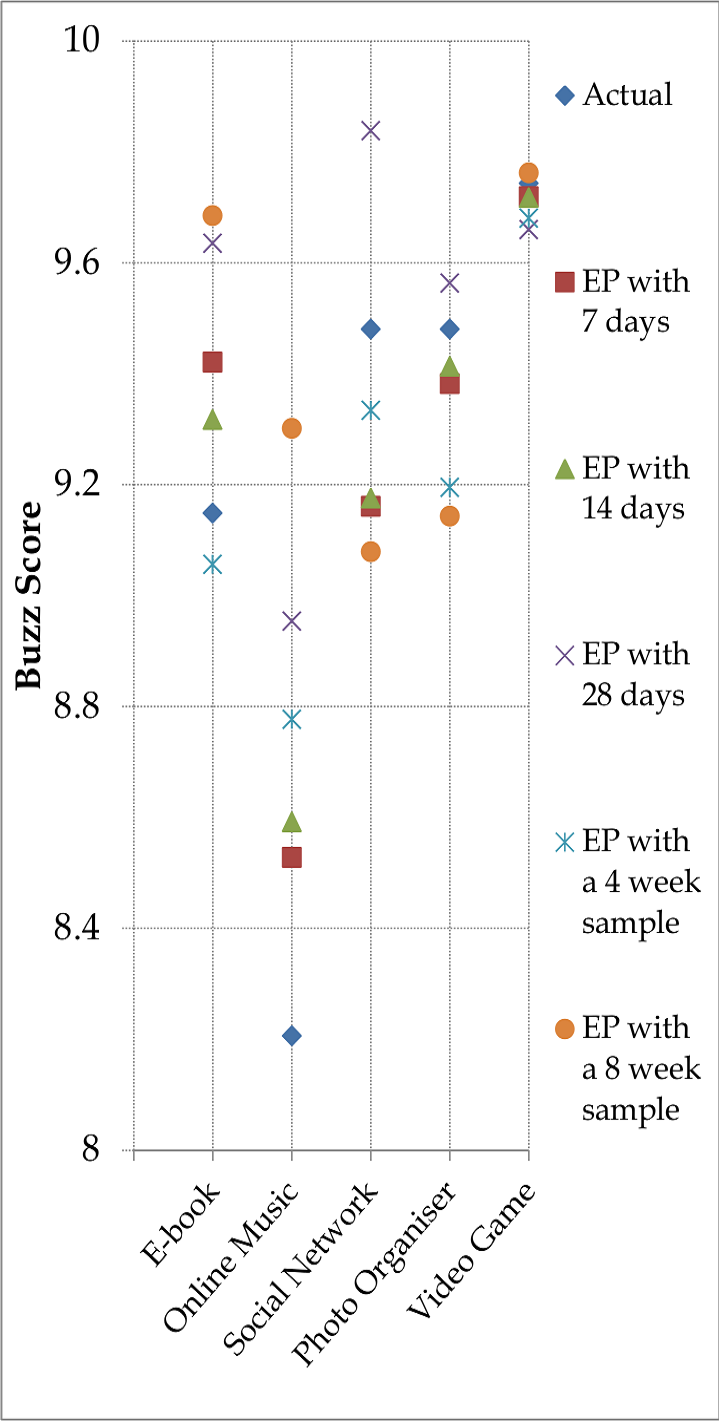}
  		\caption{EP technique}
  		\label{figure3a}
	\end{subfigure}
	\begin{subfigure} {0.49\linewidth}
		\centering
		\includegraphics[width=\linewidth]{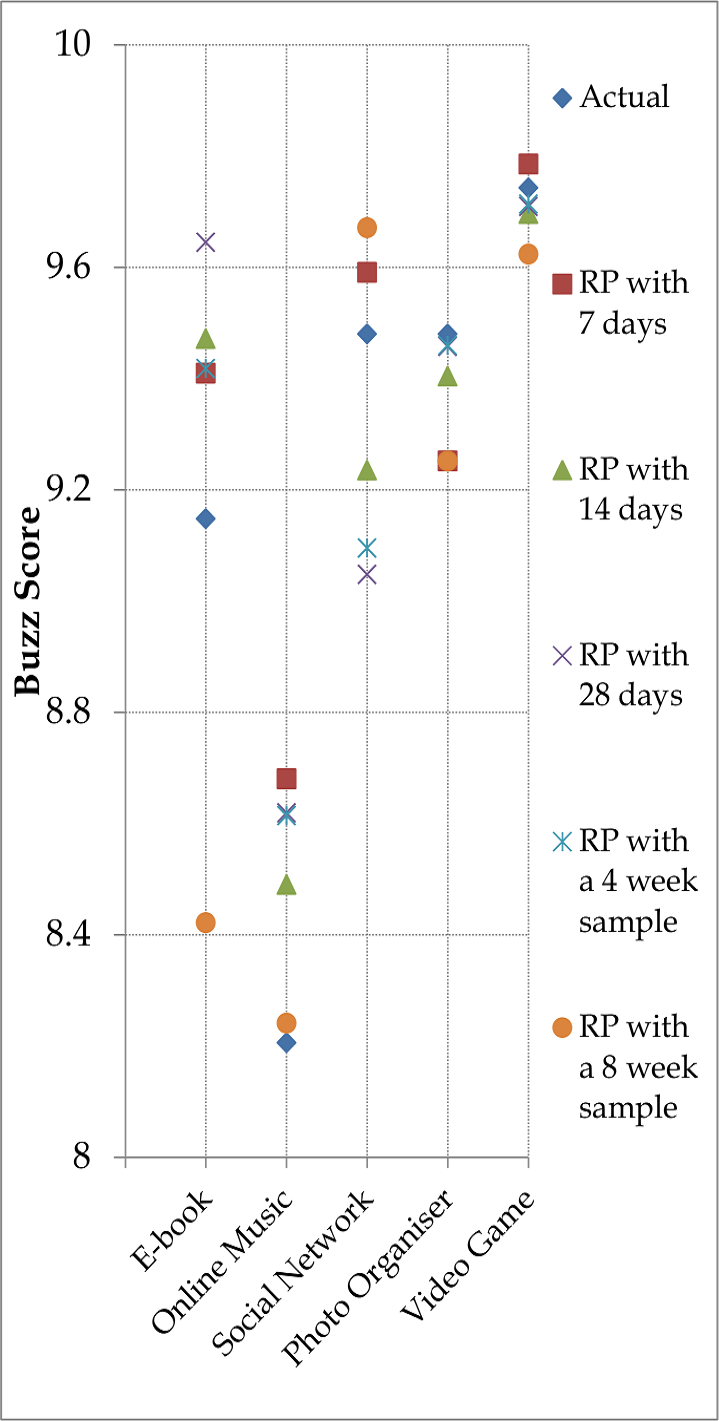}
		\caption{RP technique}
		\label{figure3b}
	\end{subfigure}%
	\caption{Predicted buzz scores for July 23, 2005 using EP and RP techniques}
	\label{figure3}
\end{figure}

The user provides as input the product (e-books), the date for which he needs to predict the buzz score (Saturday, July 23, 2005) and the $n$ value (if 14 days are selected, then July 8 to July 22 are used, or if a 4 week sample is selected, then four preceding Saturdays are used). For example, when the user selects 14 days, the \texttt{Query Constructor} takes this input and assembles the query as follows: 

\begin{verbatim}
SELECT TRANSFORM(date, buzz_score)
USING 'daily_prediction' 
FROM Yahoo_Buzz_Scores
WHERE product IN ('EBOOKS')
AND date>='2005-07-08' 
AND date <='2005-07-22'
\end{verbatim}

Figure \ref{figure3} shows the graphs plotted when the EP and RP techniques are used for predicting the buzz score of e-books, online music, social networks, photo organisers and video games on July 23, 2005. Observations made from the graph are (a) consider video game products, for which the actual buzz score is 9.74308523. The most accurate prediction using EP techniques is when an eight week sample is taken into account with an error percentage of less than 0.2\%. The most accurate prediction using RP techniques is when a four week sample is considered with less than 0.3\% error. 

Table \ref{table1} summarises the error percentage of the predicting techniques. Given the sensitivity of the buzz scores upto a 2\% error is acceptable. 

\begin{table*}
	\centering
	\begin{tabular}{| p{2cm} | p{1.1cm} | p{1.1cm} | p{1.1cm} | p{1.1cm} | p{1.1cm} |  p{1.1cm} | p{1.1cm} | p{1.1cm} | p{1.1cm} | p{1.1cm } |}
	\hline
	\multirow{2}{*}{\textbf{Technology}}	& \multicolumn{10}{c  |}{\textbf{Error (\%)}}	\\
	\cline{2-11}
	& \textbf{EP with 7 days}	& \textbf{EP with 14 days}	& \textbf{EP with 28 days}	& \textbf{EP with a 4 week sample}	& \textbf{EP with a 8 week sample}	&	\textbf{RP with 7 days}	& \textbf{RP with 14 days}	& \textbf{RP with 28 days}	& \textbf{RP with a 4 week sample}	& \textbf{RP with a 8 week sample}	\\
	\hline
	\hline	
	E-books 		&	2.98	& 1.85	& 5.32	& -1.01	& 5.86	& 2.86	& 3.53	& 5.43	& 2.95	& -7.94\\
	Online Music	&	3.92	& 4.70	& 9.11	& 6.95	& 13.34	& 5.78	& 3.46	& 5.03	& 4.97	& 0.43\\
	Social Network	&	-3.37	& -3.22	& 3.77	& -1.55	& -4.23	& 1.17	& -2.58	& -4.57	& -4.06	& 2.01\\
	Photos			&	-1.04	& -0.71	& -0.29	& -3.02	& -3.56	& -2.41	& -0.80	& -1.38	& -0.22	& -2.41\\
	Video Game		&	-0.25	& -0.27	& -0.86	& -0.65	& 0.20	& 0.44	& -0.48	& -0.34	& -0.30	& -1.22\\
	\hline	
	\end{tabular}
	\caption{Error (\%) when predicting buzz scores for e-books, online music, social network, photo organisers and video game technologies}
	\label{table1}	
\end{table*}

\subsubsection{Weekly Analysis}
\begin{figure*}
	\centering
	\begin{subfigure} {0.328\textwidth}
		\centering
  		\includegraphics[width=\textwidth]{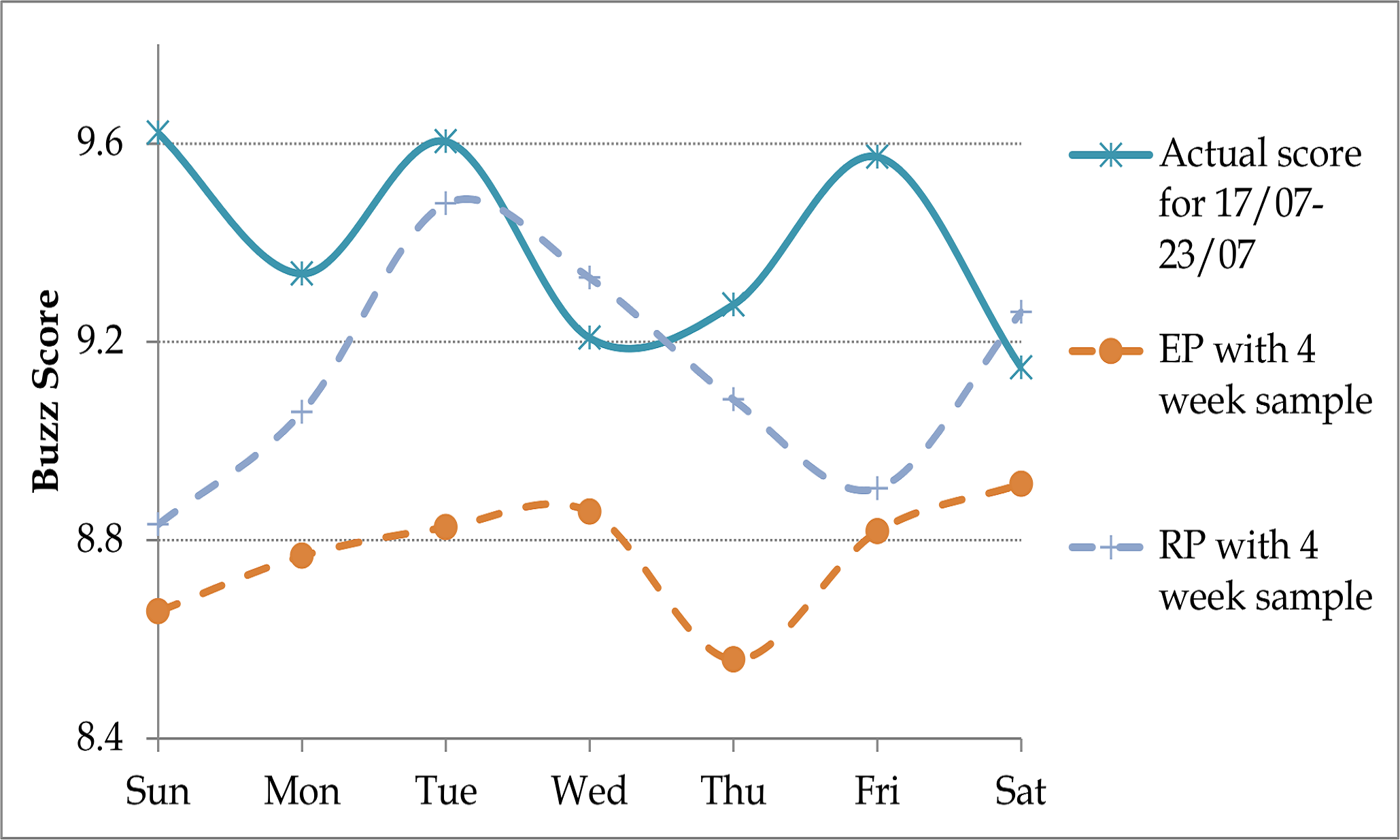}
  		\caption{E-book - 4 week sample}
  		\label{figure4a}
	\end{subfigure} \hfill
	\begin{subfigure} {0.328\textwidth}
		\centering
  		\includegraphics[width=\textwidth]{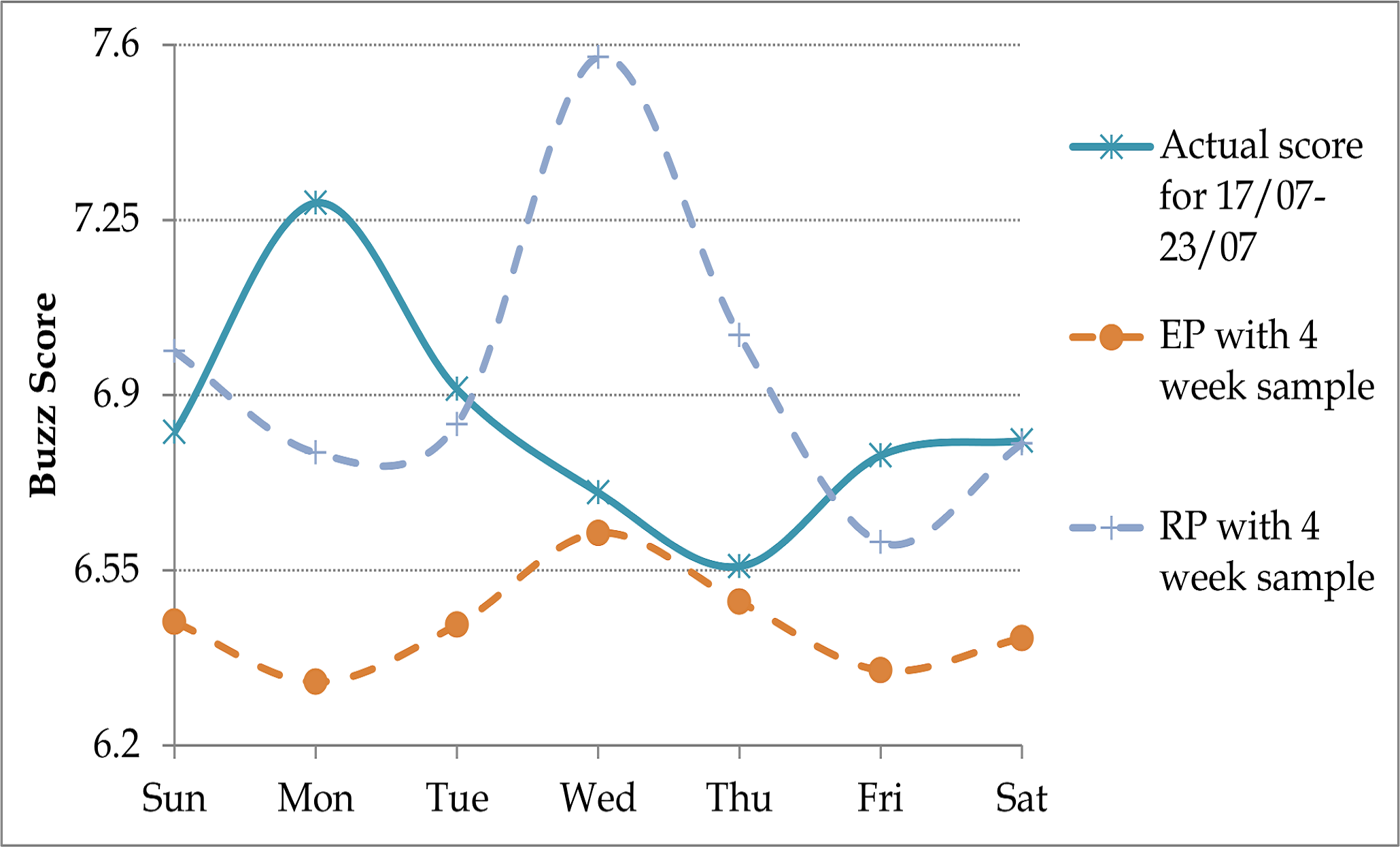}
  		\caption{Adobe Reader - 4 week sample}
  		\label{figure5a}
	\end{subfigure} \hfill
	\begin{subfigure} {0.328\textwidth}
		\centering
  		\includegraphics[width=\textwidth]{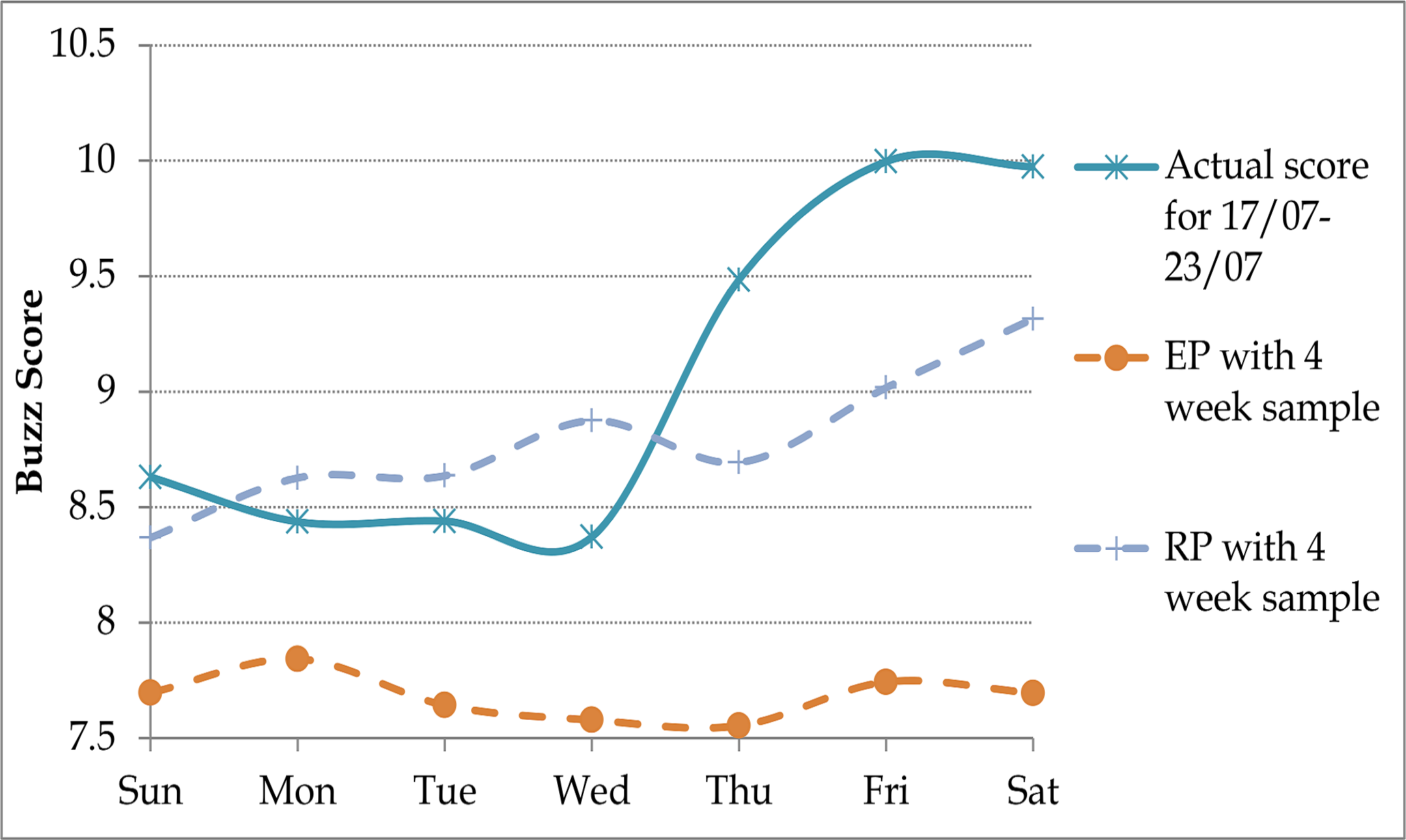}
  		\caption{Amazon - 4 week sample}
  		\label{figure7a}
	\end{subfigure}%
	\\
	\begin{subfigure} {0.328\textwidth}
		\centering
		\includegraphics[width=\textwidth]{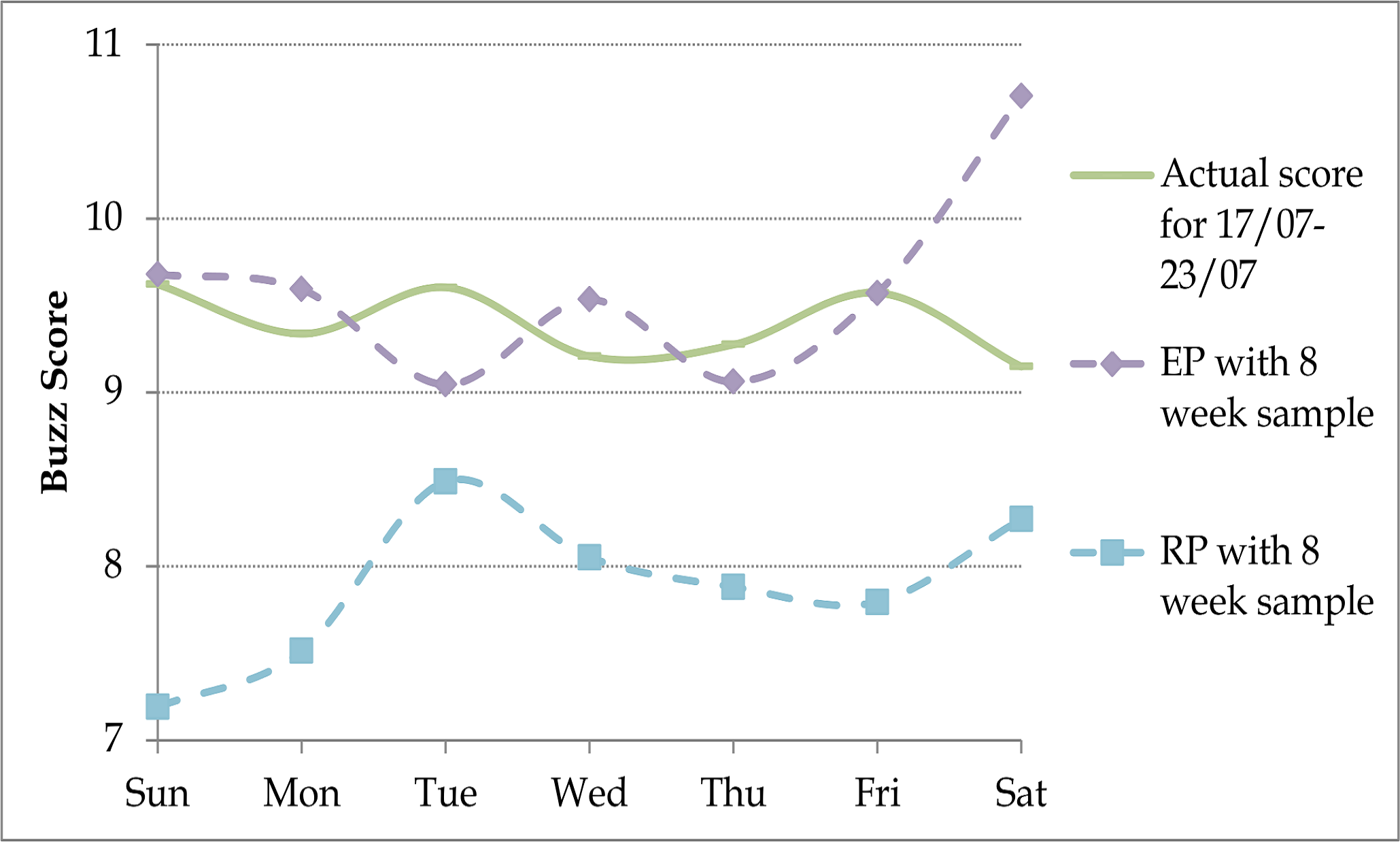}
		\caption{E-book - 8 week sample}
		\label{figure4b}
	\end{subfigure} \hfill
	\begin{subfigure} {0.328\textwidth}
		\centering
		\includegraphics[width=\textwidth]{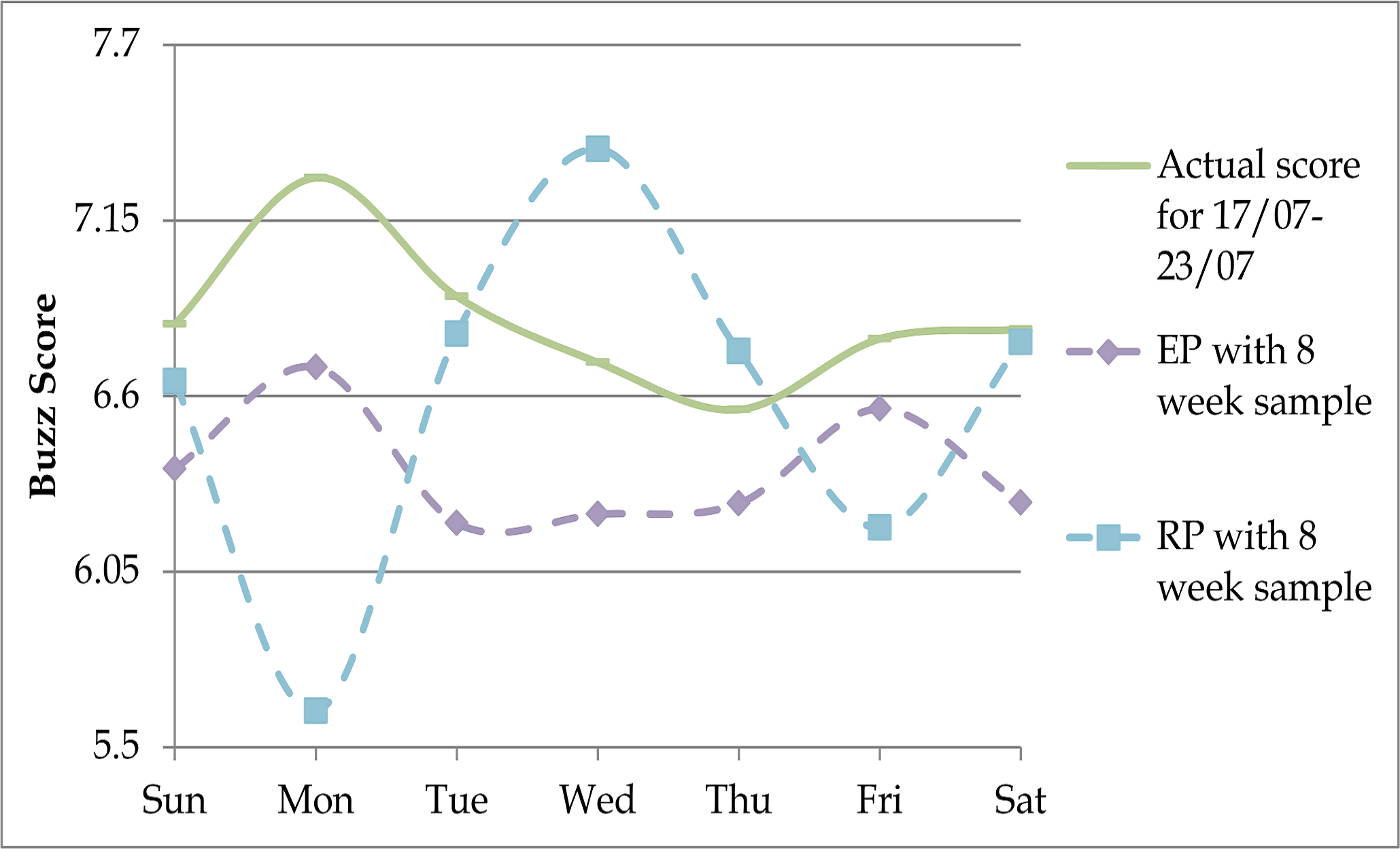}
		\caption{Adobe Reader - 8 week sample}
		\label{figure5b}
	\end{subfigure} \hfill
	\begin{subfigure} {0.328\textwidth}
		\centering
  		\includegraphics[width=\textwidth]{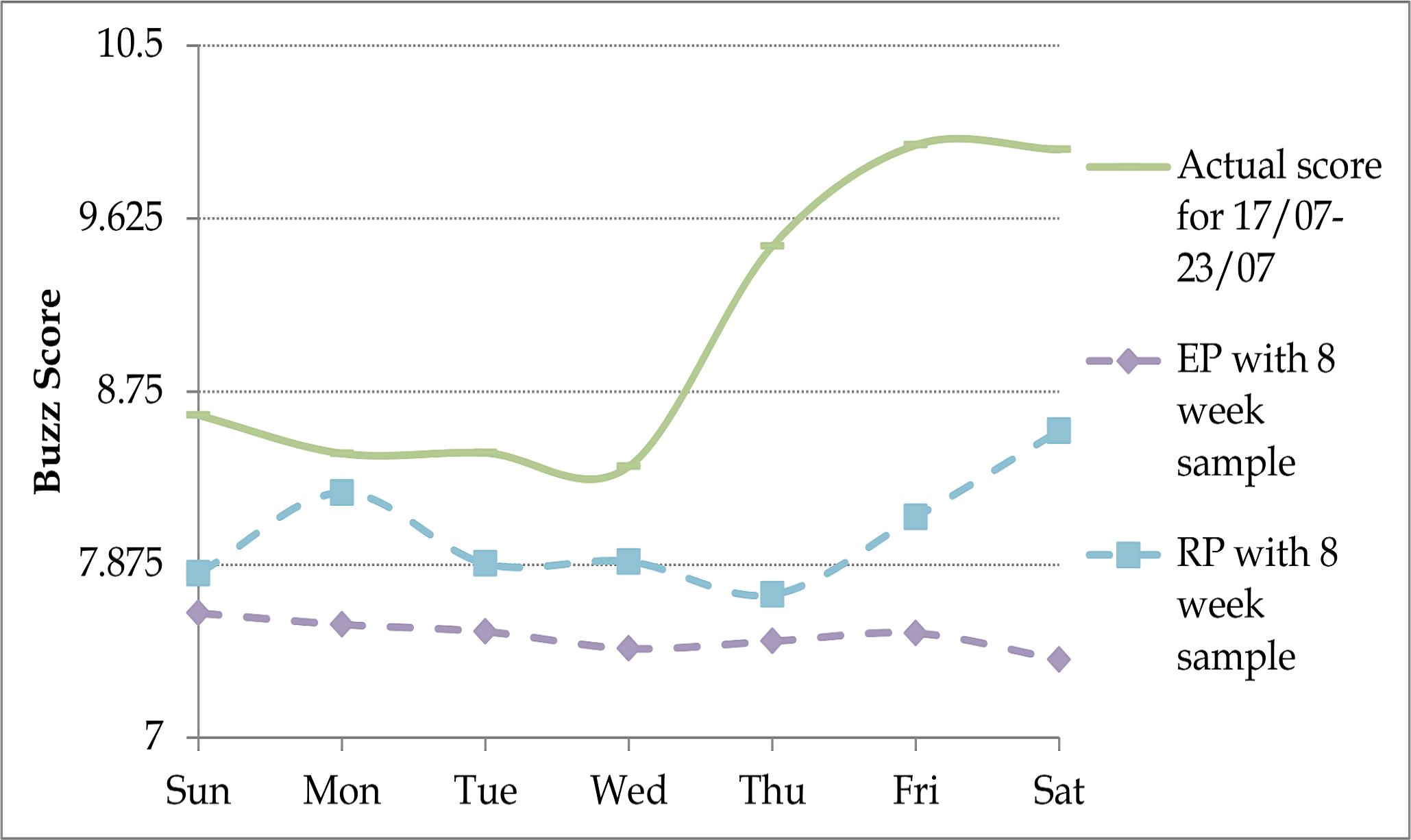}
  		\caption{Amazon - 8 week sample}
  		\label{figure7b}
	\end{subfigure}%
	\caption{Predicted buzz scores of E-books (including Adobe Reader and Amazon) for July 17 - 23 using EP and RP techniques}
	\label{figure4}
\end{figure*}

The techniques employed for predicting daily buzz scores were applied for predicting weekly trends. Figure \ref{figure4} shows the buzz scores for e-books, including Adobe and Amazon e-books, for July 17-23, 2005 using both the EP and RP techniques. A sample of any preceding four weeks and 8 weeks are used in the prediction techniques. The graphs indicate that the trends are better captured by the RP technique when four and eight week samples are used. As shown in Table \ref{table2} a correlation of over 70\% (lowest is just less than 30\%) can be obtained when RP techniques are used. The correlation between the actual and predicted buzz scores are as low as 1\% (highest is 68\%) when the EP technique is employed. 

\begin{table}
	\centering
	\begin{tabular}{| p{1.7cm} | p{1.2cm} | p{1.2cm} | p{1.2cm} | p{1.2cm} | }
	\hline
	\multirow{2}{*}{\textbf{Technology}}	& \multicolumn{4}{c  |}{\textbf{Correlation}}	\\
	\cline{2-5}
	&	EP with 4 samples	&	RP with 4 samples	&	EP with 8 samples	&	RP with 8 samples\\
	\hline
	\hline
	E-book		&	-0.22	&	-0.34	&	-0.45	&	-0.29\\
	- Adobe Book	&	-0.62	&	-0.38	&	0.68	&	-0.73\\
	- Amazon	&	0.01	&	0.70	&	-0.47	&	0.44\\
	\hline
	\end{tabular}
	\caption{Correlation between actual and predicted buzz scores for e-books, and for Adobe and Amazon e-books}
	\label{table2}	
\end{table}

\subsection{Case Study 2: Qualitative Interpretation of News Related $n$-grams}

The `Yahoo! n-Grams' \footnote{\url{http://webscope.sandbox.yahoo.com/catalog.php?datatype=l}}, version 2.0, is used in this case study. This dataset is based on data gathered from over 12,000 news related sites, 14.6 million documents, 126 million unique sentences, 3.4 billion running words between February 2006 and December 2006. The data set contains $n$-grams (continuous words, where $n = 1$ to $5$) and their frequency. 

The uncompressed data is 40 GB and the data related to 5-grams is 12 GB. We use this data in the BigExcel framework for searching the frequencies of certain key words related to important events and qualitatively interpreting the results. We use the 5-gram dataset for this case study and three events that occurred in 2006.  

The first event is the pandemic threat of H5N1 virus, otherwise referred to as bird flu. This event evolved throughout the year with the first sighting in early January in Turkey resulting in a number of deaths, followed by similar outbreaks in a numerous African, Asian and European countries leading to large number of deaths throughout the year \cite{h5n1-1}. Given the widespread nature of the event and its severity we expect that news websites may have given significant importance to this event. 

The second event considered is the 2006 World Cup football tournament which happened in June and July. Italy won the tournament by beating France. Over the course of the tournament nearly 26.29 billion non-unique users watched matches and obtained the largest number of television views\footnote{\url{http://www.fifa.com/mm/document/fifafacts/ffprojects/ip-401_06e_tv_2658.pdf}}. We expect that this event would have received significant coverage on news websites.    

The third event is the purchase of Youtube by Google on October 9, 2006\footnote{\url{http://www.nytimes.com/2006/10/09/business/09cnd-deal.html}}. Google beat a number of competitors including Microsoft and Yahoo in its \$1.65 billion purchase. We expect this news item would have been featured on news sites. 

In the dataset there are 29,570,136 5-grams and we explore the number of 5-grams that were related to the three events. For example, consider the bird flu event. We explore the dataset to determine whether there are 5-grams related to the event by searching for relevant key words. 

\begin{figure*}
	\centering
	\begin{subfigure} {0.48\textwidth}
		\centering
  		\includegraphics[width=\textwidth]{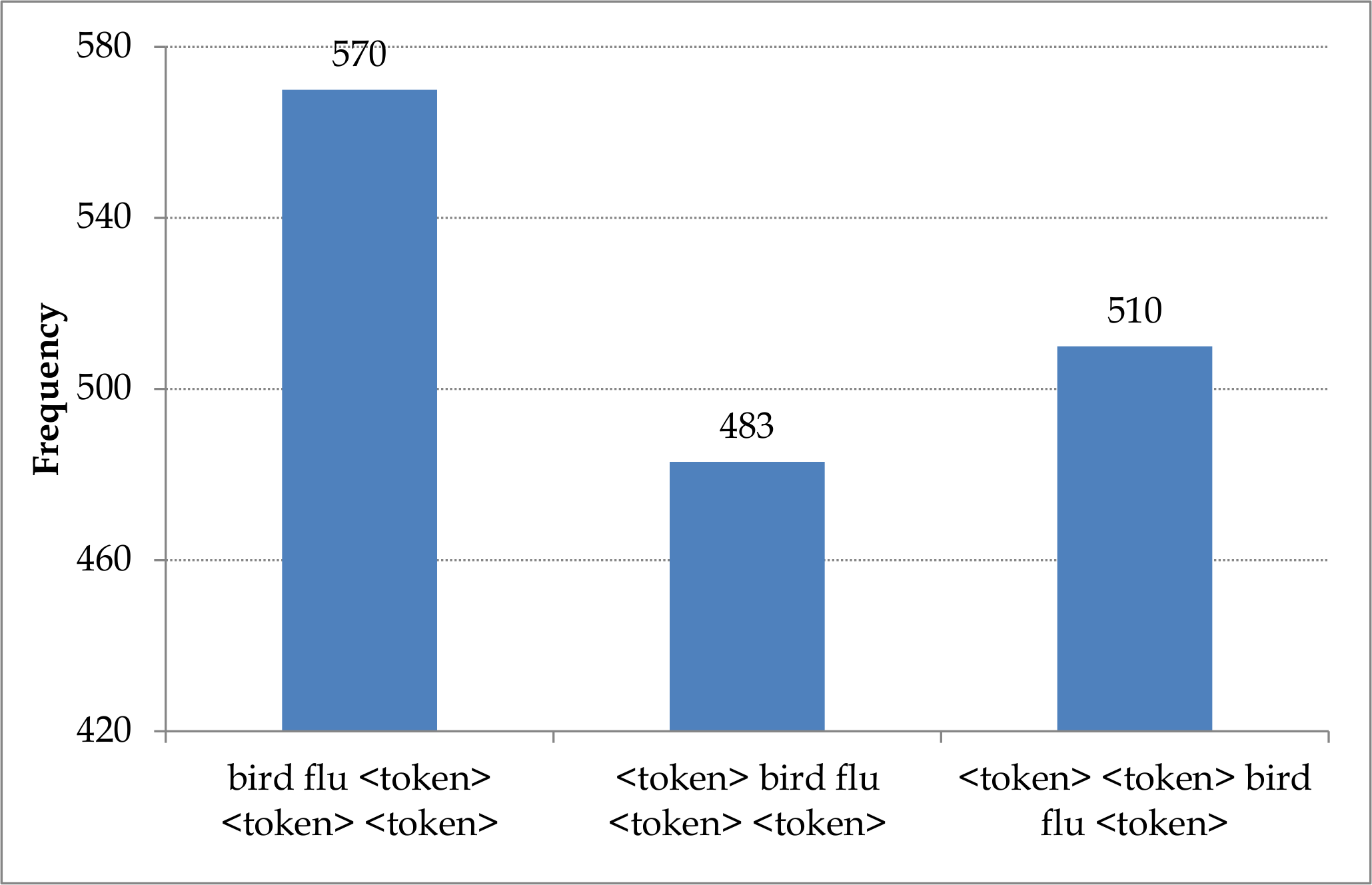}
  		\caption{Bird flu event}
  		\label{figure8a}
	\end{subfigure} \hfill
	\begin{subfigure} {0.48\textwidth}
		\centering
		\includegraphics[width=\textwidth]{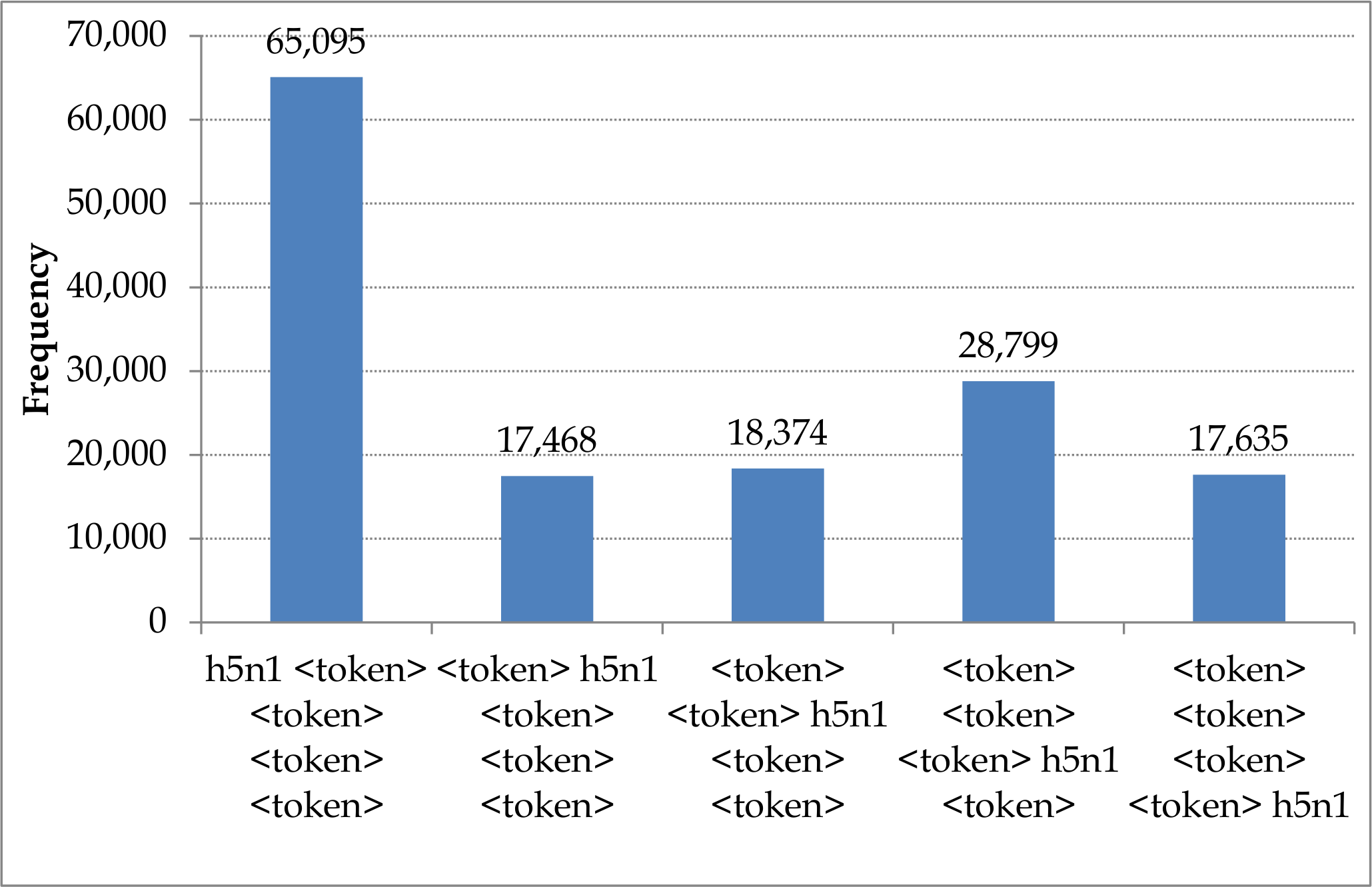}
		\caption{Bird flu event (contd.)}
		\label{figure8a1}
	\end{subfigure} \\
	\begin{subfigure} {0.48\textwidth}
		\centering
  		\includegraphics[width=\textwidth]{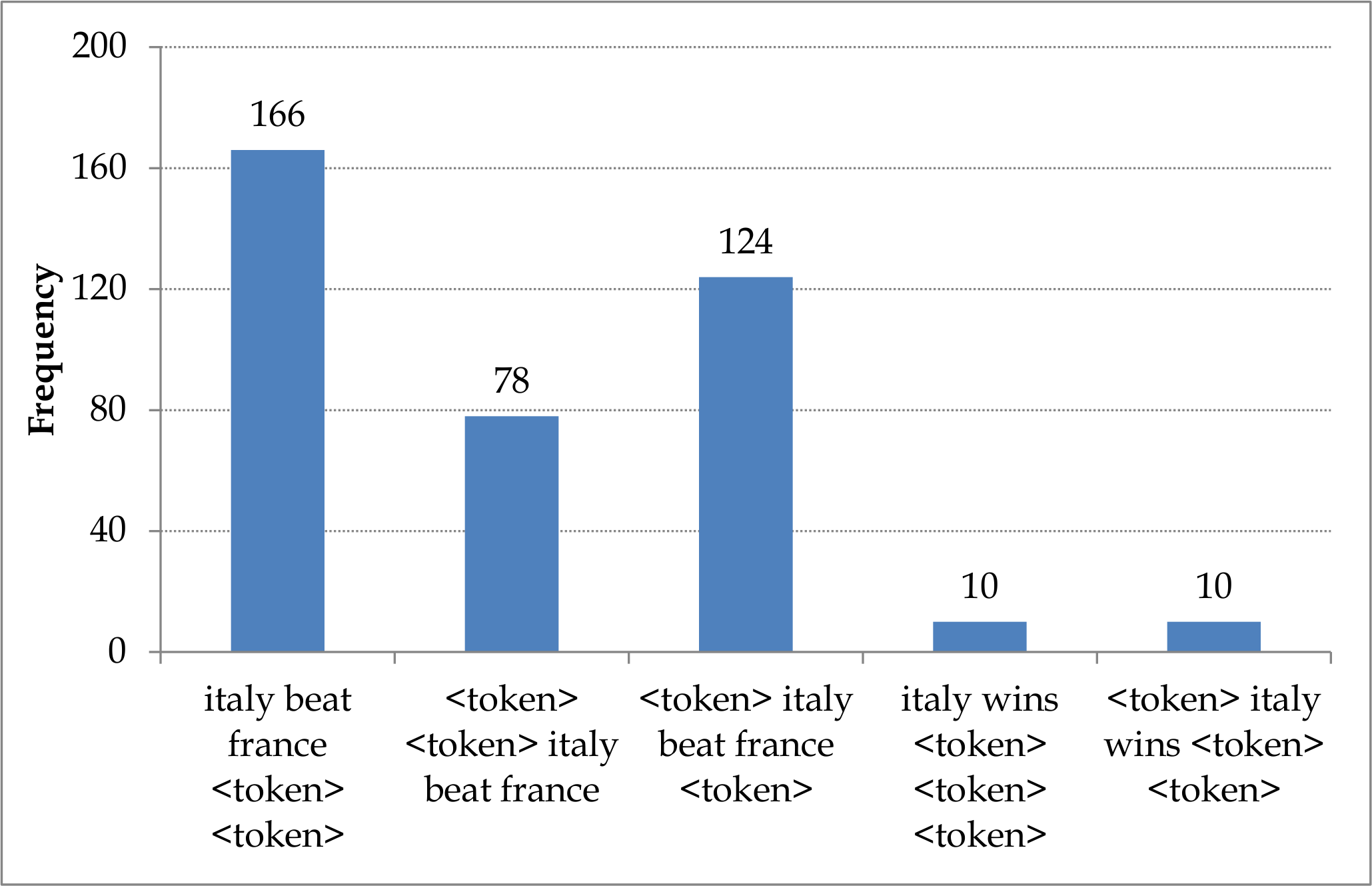}
  		\caption{World Cup football}
  		\label{figure8b}
	\end{subfigure} \hfill
	\begin{subfigure} {0.48\textwidth}
		\centering
		\includegraphics[width=\textwidth]{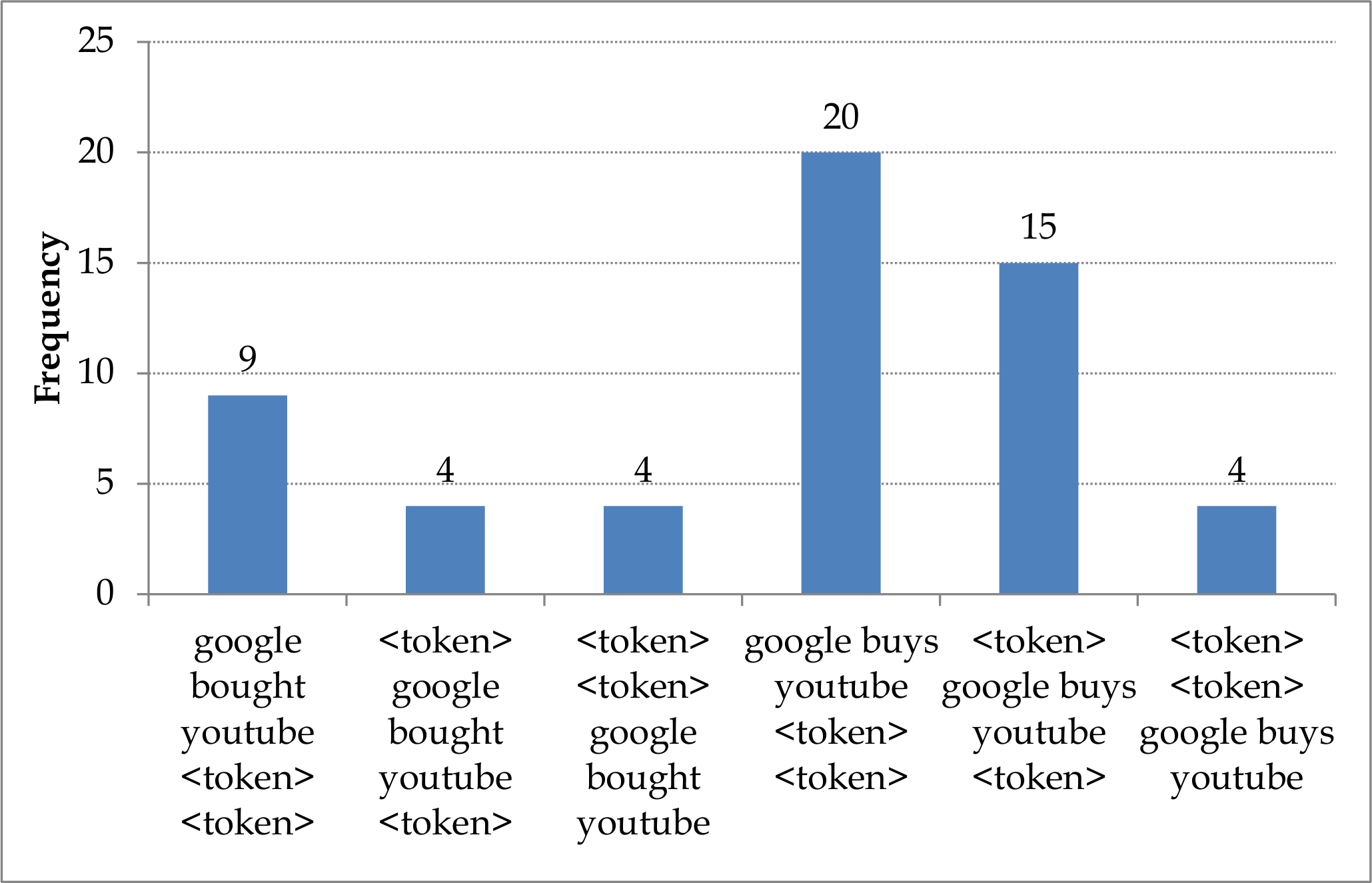}
		\caption{Google buys Youtube}
		\label{figure8c}
	\end{subfigure} 
	\caption{Frequency of 5-grams related to three important events in 2006}
	\label{figure8}
\end{figure*}

Figure \ref{figure8a} and Figure \ref{figure8a1} shows the frequency of the 5-grams related to the bird flu. The 2-gram, bird flu, is indicated while all the other tokens that surround the 2-gram are denoted using $<token>$. For example, `$<token>$ $<token>$ $bird$ $flu$ $<token>$' could be `$tenth$ $human$ $bird$ $flu$ $death$' or '$more$ $human$ $bird$ $flu$ $death$'. There are 148,934 5-grams directly related to the event which is 0.5037\% of the entire corpus.

To perform the above, the user provides as input the table name (\texttt{Yahoo\_n-grams}), column names (\texttt{n-gram}, \texttt{frequency} and the name of the module that for analysis (\texttt{event\_analysis}) through the data browser. The \texttt{Query Constructor} then assembles the following query:

\begin{verbatim}
SELECT TRANSFORM(n-gram, frequency)
USING 'ngram_analysis' 
AS distinct_n-gram, total_frequency
FROM Yahoo_n-grams
\end{verbatim}

The \texttt{ngram\_analysis} module is embedded from the \texttt{Module Repository} into the query by the \texttt{Analytics Processor}. The module includes the logic for searching the corpus for tokens related to the bird blu and then organising the data into various distinct ngrams and their frequencies. For example, the distinct n-grams related to 'bird flu' are $bird$ $flu$ $<token>$ $<token>$ $<token>$, $<token>$ $bird$ $flu$ $<token>$ $<token>$, $<token>$ $<token>$ $bird$ $flu$ $<token>$, and $<token>$ $<token>$ $<token>$ $bird$ $flu$. The output is parsed by the \texttt{Response Processor} to display the information in a meaningful way, for example, to generate data that is represented in Figure \ref{figure8a} and Figure \ref{figure8a1}. 

Figure \ref{figure8b} is the frequency of the 5-grams related to the world cup football event. The 5-grams related to the event are based on phrases for example, `italy beat france' and 'italy wins'. There are 388 5-grams directly related to the event which is approximately 0.0013\% of all the 5-grams contained in the corpus. 

Figure \ref{figure8c} is the frequency of the 5-grams related to the purchase of Youtube by Google. There are 56 5-grams related to the event.  
 
There are about four hundred times more 5-grams directly related to the bird flu, an event which evolved during the whole year, than the finals of the world cup football match. Similarly, the world cup final has seven times more coverage than the takeover of Youtube by Google. Using this analysis it may not be possible to make firm conclusions but it would seem that the bird flu generated nearly 30 times more news each week compared to the weeks when the world cup football was going on. 

\section{Discussion and Conclusions}
\label{conclusions}

This research identified three big data challenges, namely (i) the accessibility of big data technologies for non-computer scientists, (ii) the ad hoc exploration of large data sets with minimal efforts and (iii) the availability of lightweight frameworks for ad hoc analytics on the web. We developed the BigExcel framework to address the above challenges. The framework is three tiered and web-based that facilitates the management of user interactions with large data sets, the construction of queries to explore the data set and the management of the infrastructure. The feasibility of BigExcel was demonstrated using two Yahoo Sandbox datasets in which quantitative and qualitative analysis was performed. A video demonstration of the framework with the source code is available at \url{http://bigdata.cs.st-andrews.ac.uk/projects/bigexcel-exploring-big-data-for-social-sciences/}. 

In the future, we aim to extend BigExcel to accommodate the analysis of unstructured data for fully exploiting the benefits of big data technologies \cite{bigdata-2, bigdata-3}. Additionally, efforts need to be directed towards automating data management by taking minimum input from the user. This will make the framework more accessible to non-Computer science communities. Another direction for enhancing BigExcel is developing generic algorithms for the \texttt{Module Repository} and here machine learning approaches will need to be incorporated for dealing with numerous types of data and their analyses \cite{bigdata-2}. Further, the repository can benefit from the development and integration of analytical libraries that support mathematical and statistical functions.

\section{Acknowledgements}

This research was pursued through an Amazon Web Services Education Research Grant and was facilitated by the availability of the Yahoo Sandbox datasets. The first author was the recipient of an Erasmus Mundus scholarship.

\end{document}